\documentstyle[11pt,paspconf,epsf]{article}

\def\etal{{\it et al.\ }}
\def\lsim{\raise0.3ex\hbox{$<$}\kern-0.75em{\lower0.65ex\hbox{$\sim$}}} 
\def\gsim{\raise0.3ex\hbox{$>$}\kern-0.75em{\lower0.65ex\hbox{$\sim$}}} 

\begin{document}

\title{Numerical Simulations of Rotating Accretion Flows near a Black Hole}
\author{Dongsu Ryu\altaffilmark{1}}
\affil{Dept. of Astronomy, Univ. of Washington, Seattle, WA 98195-1580\\
Dept. of Astronomy \& Space Science, Chungnam Nat. Univ., Korea}
\author{Sandip K. Chakrabarti\altaffilmark{2}}
\affil{S.N. Bose Center for Basic Sciences, Calcutta 700091, India}

\altaffiltext{1}{email: ryu@hermes.astro.washington.edu}
\altaffiltext{2}{email: chakraba@bose.ernet.in}

\begin{abstract}

We present time-dependent solutions of thin, supersonic accretion
flows near a black hole and compare them with analytical solutions.
Such flows of inviscid, adiabatic gas are characterized by the specific
angular momentum and the specific energy.
We confirm that for a wide range of above parameters a stable standing
shock wave with a vortex inside it forms close to the black hole.
Apart from steady state solutions, we show the existence of non-steady
solutions for thin accretion flows where the accretion shock is
destroyed and re-generated periodically.
The unstable behavior should be caused by dynamically induced
instabilities, since inviscid, adiabatic gas is considered.
We discuss possible relevance of the periodic behavior on
quasi-periodic oscillations (QPOs) observed in galactic and extragalactic
black hole candidates.

\end{abstract}

\section{Introduction}

Rotating accretion flows are important ingredients in many astrophysical
systems containing a black hole, which involve mass transfer from one
object to another (such as in a binary system) or from set of objects
to another (such as in a galactic center).
The standard disk model of such accretion flows by Shakura \& Sunyaev
(1973) assumes Keplerian distribution of accreting matter.
There, the inner edge of the disk is chosen to coincide with the marginally
stable orbit located at three Schwarzschild radii, $r_i=3R_g$, where
the Schwarzschild radius, $R_g=2GM_{BH}/c^2$, is the horizon of a black
hole of mass $M_{BH}$.
This disk model is clearly incomplete, since the inner boundary condition
on the horizon was not taken care of.
As an accretion flow approaches the horizon, its radial velocity reaches the
velocity of light.
Therefore, a black hole accretion flow is necessarily supersonic and must
pass through a sonic point where the flow has to be sub-Keplerian.
Thus, independent of heating and cooling processes, a black hole accretion
has to deviate from a standard Shakura-Sunyaev type Keplerian disk.
The disk with realistic accretion flows is called the advective disk
(Chakrabarti 1996).
Here, we present the results of numerical study of accretion flows
near a black hole by assuming they are thin, axisymmetric, and inviscid.

\section{Analytic Consideration}

We choose cylindrical coordinates $(r,\theta, z)$ and place a black hole
at the center.
We assume that the gravitational field of the black hole can be described
in terms of the potential introduced by Paczy\'nski \& Wiita (1980)
\begin{equation}
\phi(r,z) = -{GM_{bh}\over R-R_g}_,
\end{equation}
where $R=\sqrt{r^2+z^2}$.
The accreting matter is assumed to be adiabatic gas without cooling and
dissipation and described with a polytropic equation of state,
$P=K \rho^{\gamma}$, where $\gamma$ is the adiabatic index which is
considered to be constant with $4/3$ throughout the flow.
$K$ is related to the specific entropy of the flow, $s$, and varies
only at shocks, if present.
Since we consider only weak viscosity limit, the specific angular momentum
of the accretion flow, $\lambda=rv_{\theta}$, is assumed to be conserved. 
Thus, unlike a Bondi flow, which is described by a single parameter
(say, specific energy), the one-dimensional accretion flows are 
described by two parameters which are the specific energy, 
${\cal E}$, and the specific angular momentum, $\lambda$.
Fig. 1 shows the classification in the parameter space (Chakrabarti 1989;
Ryu \& Chakrabarti 1997):\hfill\break
N:     No sonic points.
       Shock only if supersonic injection.\hfill\break
O:     Outer sonic point only as in a Bondi solution.
       No shock.\hfill\break
I:     Inner sonic point only.
       Shock only if supersonic injection.\hfill\break
O$^*$: Outer and center sonic points.
       Solution does not extend to the horizon.\hfill\break
I$^*$: Inner and center sonic points.
       Solution does not extend to large distance.\hfill\break
SA:    Two (outer and inner) sonic points.
       Shock in accretion solutions but not in wind solutions.\hfill\break
SW:    Two sonic points.
       Shock in wind solutions but not in accretion solutions.\hfill\break
NSA:   Two sonic points.
       No shock condition satisfied in accretion solutions.\hfill\break
NSW:   Two sonic points.
       No shock condition satisfied in wind solutions.\hfill\break

\begin{figure}
\vspace{0truein}
\epsfysize=3.3in\epsfbox[-10 160 530 580]{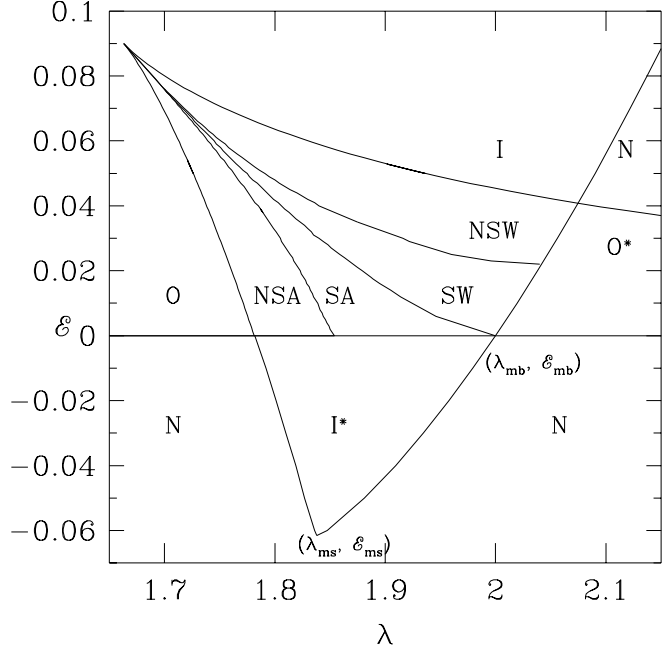}
\vspace{0truein}
\caption{Classification of one-dimensional accretion flows in the parameter
space of specific energy, ${\cal E}$, and the specific angular momentum,
$\lambda$.
${\cal E}$ is given in unit of $c^2$ and $\lambda$ in unit of $R_gc$.}
\vspace{0truein}
\end{figure}

Although these solutions are strictly valid for inviscid flows, even
when viscosity is high, given that the viscous time-scale is likely to be
much larger compared to the infall time-scale, the inviscid solutions
are likely to remain important.

\section{One-dimensional Numerical Solutions}

Fig. 2 shows an example numerical solution from the `SA' region.
We superposed analytical solution (solid lines) and numerical solutions
with TVD code (dashed lines) and with SPH code (dotted lines)
(Molteni, Ryu \& Chakrabarti 1996).
The TVD calculation was done with $512$ grids and the SPH calculation
was done with $\sim 560$ particles of size $h=0.3R_g$.
Matter was injected at the outer boundary located at $x=50R_g$.
An absorption condition was used to mimic the black hole horizon at
the inner boundary which is located either at $x=1.5R_g$ for the TVD
calculation or at $x=1.25R_g$ for the SPH calculation.
The adiabatic index $\gamma=4/3$ was used.

\begin{figure}
\vspace{0truein}
\epsfysize=2.95in\epsfbox[-150 90 500 645]{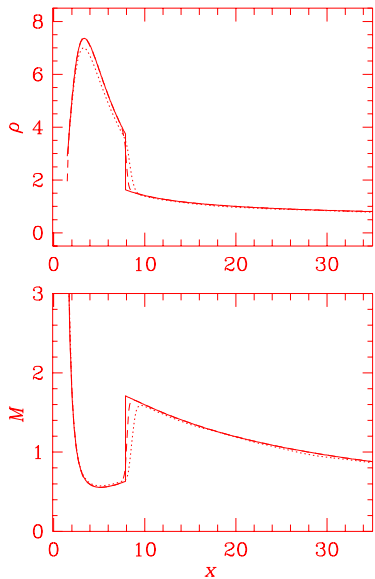}
\vspace{0truein}
\caption{One-dimensional accretion flow with a standing shock in the
region of `SA'.
${\cal E}=0.036$ and $\lambda=1.80$ are used.
The solid line is the analytical solution, and the long and short dashed
lines are the solutions of the TVD and SPH simulations, respectively.
Upper panel is the mass density in arbitrary units and the lower panel
is the Mach number of the flow.}
\vspace{0truein}
\end{figure}

The figure shows an excellent agreement between the analytic and
numerical solutions.
Here, the flow starts out subsonically, presumably from a Keplerian disk.
Then, it enters through the outer sonic point (located at $x=27.97R_g$),
passes through the shock (located at $x=7.98R_g$), and subsequently passes
through the inner sonic point (located at $x=2.57R_g$) before entering
the black hole.

\section{Two-dimensional Numerical Solutions}

In multi-dimensional accretion flows, non-steady solutions, as well as
steady solutions, exist.
Here, we discuss some examples of accretion flows with zero specific
energy $({\cal E}=0)$.
More extensive discussion on accretion flows with ${\cal E}=0$ was
reported in Ryu, Chakrabarti \& Molteni (1997).
Discussion on extensive calculations for accretion flows with non-zero
specific energy $({\cal E}\ne0)$ will be reported elsewhere
(Ryu \& Chakrabarti 1997).

In these simulations, we inject supersonic 
matter (with a radial Mach number $M=v_r/a=10$) 
at the outer boundary, $r_b=50R_g$.
The inflow at the outer boundary is assumed to have a small thickness,
$h_{in}$, or a small arc angle, $\theta_{in}=\arctan(h_{in}/r_b)\ll1$.
If zero-energy accretion flows belong the region of `O', most of the
material is accreted into the black hole forming a stable quasi-spherical
flow or a simple disk-like structure around it (just as a Bondi flow).
If accretion flows belong the region of `SA' and `O$^*$', the incoming
material produces a stable standing shock with one or more vortices behind
it and some of it is deflected away at the shock as a conical outgoing
wind of higher entropy.

Accretion flows with parameters in the region `NSA' with
$1.782R_gc<\lambda<1.854R_gc$ show an unstable behavior.
Fig. 3 shows an example numerical solution with $\lambda=1.85R_gc$.
In this case, the structure with an accretion shock and a generally
subsonic high density disk is established around the black hole.
However, the structure is not stable.
At the accretion shock, the incoming flow is deflected.
But some of the post-shock flow, which is further accelerated by the
pressure gradient behind the shock, goes through a second shock,
where the flow is deflected once more downwards.
The downward flow squeezes the incoming material, and the
accretion shock starts collapsing $(t=1.4\times 10^4R_g/c)$.
In the process of the collapse, some of the post-shock material escapes
as wind but most is absorbed into the black hole.
After the collapse, the re-building of the accretion shock starts with
the incoming material bouncing back from the centrifugal barrier.
The subsonic post-shock region becomes a reservoir of material,
so the material is accumulated behind the shock.
With the accumulated material a giant vortex is formed, which in turn
supports the accretion shock $(t=2\times 10^4R_g/c)$.
This continues until the incoming flow is squeezed enough so the accretion
shock collapsed, and the cycle continues.

\begin{figure}
\vspace{0truein}
\epsfysize=6.in\epsfbox[25 60 600 700]{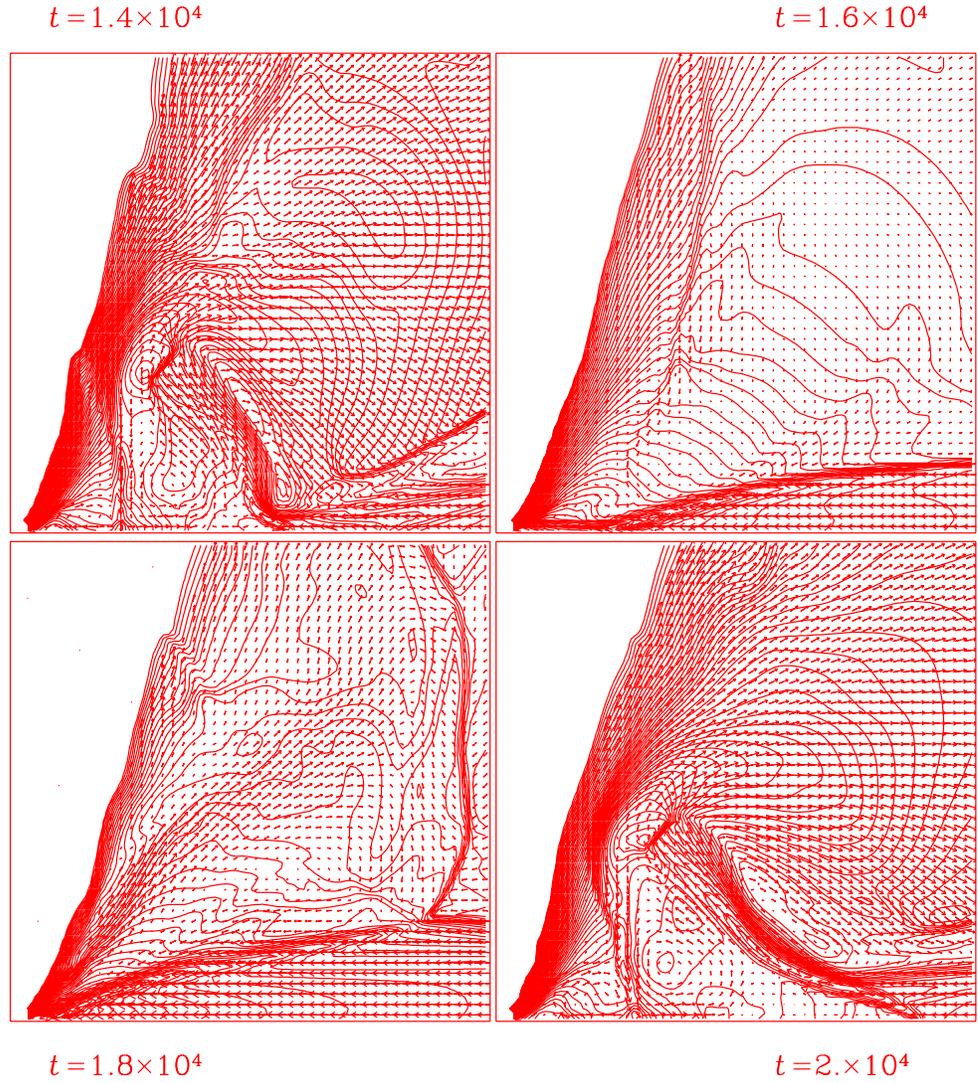}
\vspace{0truein}
\caption{Two-dimensional accretion flow showing an unstable behavior in
the region of `NSA'.
${\cal E}=0$ and $\lambda=1.85$ are used.
Contours are for density and arrows are for velocity vector.
Time is given in unit of $R_g/c$.}
\vspace{0truein}
\end{figure}

\section{Discussion}

The time scale of the periodicity of unstable flows is interesting.
It is in the range of
\begin{equation}
\tau \approx 4-6\times10^3{R_g\over c}=4-6\times10^{-2}
\left({M_{bh}\over M_{\sun}}\right){\rm s}.
\end{equation}
The modulation of amplitude is also very significant, and could be
as much as a hundred percent depending on detailed processes.
Oscillations with these characteristics have been observed in black hole
candidates and are called the QPOs.
For instance, in the QPOs from the low mass x-ray binaries, the
oscillation frequency has been found to lie typically between 5 and 60 Hz
(Van der Klis 1989).
Thus, compact objects with mass $M \approx 0.3-5M_\odot$ could 
generate oscillations of right frequencies due to the instability
discussed in this paper. Similar oscillations of period on the
order of a few hours to a few days are expected in soft X-rays and UV
emissions from galactic centers.
However, by considering simplified physics which we have assumed here,
it may be premature to assume that the presented mechanism would explain
all the QPOs observed. In some of the cases, the oscillation may
be due to the dynamic instability considered in Ryu \etal (1995),
or it may be due to resonance of the cooling time scale (bremsstrahlung
or Comptonization, whatever the case may be) and the infall time scale
in the enhanced density region near the centrifugal barrier as shown
by Molteni, Sponholz and Chakrabarti (1996).
In detailed works, radiative processes as well as viscosity should be
included in the accretion calculations to examine the observational
consequences of the present instability.

\acknowledgments

The work by DR was supported in part by Seoam Scholarship Foundation.

\end{document}